\documentclass[nature,twocolumn,preprintnumbers,amsmath,amssymb,superscriptaddress]{revtex4}

\usepackage{graphicx}
\usepackage{dcolumn}
\usepackage{subfigure}

\newcommand{\be}{\begin{equation}}
\newcommand{\ee}{\end{equation}}
\newcommand{\bea}{\begin{eqnarray}}
\newcommand{\eea}{\end{eqnarray}}
\newcommand{\bm}[1]{\mathbf{#1}}
\newcommand{\la}{\langle}
\newcommand{\ra}{\rangle}

\newcommand{\lp}{\left(}
\newcommand{\rp}{\right)}
\newcommand{\ty}[1]{\mbox{\tiny #1}}

\def \cF{{\cal F}}

\def \cA{{\cal A}}

\def \cT{{\cal T}}

\begin{document}

\title{Transport Between Twisted Graphene Layers}

\author{R. Bistritzer and A.H. MacDonald}
\affiliation{Department of Physics, The University of Texas at Austin, Austin Texas 78712\\}

\date{\today}

\begin{abstract}
Commensurate-incommensurate transitions are ubiquitous in physics and are often accompanied by intriguing phenomena.
In few-layer graphene (FLG) systems, commensurability between honeycomb lattices on adjacent layers
is regulated by their relative orientation angle $\theta$, which is in turn dependent on sample preparation procedures.
Because incommensurability suppresses inter-layer hybridization, it is often claimed that graphene
layers can be electrically isolated by a relative twist, even though they are vertically separated by a fraction of a nanometer.
We present a theory of interlayer transport in FLG systems which reveals a richer picture in which the specific conductance
depends sensitively on $\theta$, single-layer Bloch state lifetime, in-plane magnetic field, and
bias voltage.  We find that linear and differential conductances are generally large and negative near commensurate values of $\theta$,
and small and positive otherwise.
\end{abstract}

\maketitle

Experimental advances in the fabrication of graphene-based structures\cite{review1,review2} have now provided
researchers with a multitude of systems that have strikingly distinct electronic properties.
By engineering the substrate underlying exfoliated samples \cite{SiO2,mica,ruthenium},
identifying exfoliated fragments with folds\cite{Haug}, or
controlling epitaxial growth conditions\cite{deheer,first},
the size and shape of the
honeycomb lattice arrays \cite{largeGraphene,nanoribbon} and
the number of graphene layers and their orientations can all be varied.
This structural diversity nourishes hopes for a future carbon-based electronics\cite{carbonElectronics}
with band-structure and transport characteristics that can be tailored for different types of applications.

FLG has advantages over single-layer-graphene because it has a larger current-carrying capacity
and because its electronic properties are sensitive to more engineerable system parameters\cite{min2008}.
In nature it appears in a variety of stacking arrangements,
the most common being Bernal and rhombohedral sequences which can form three dimensional
lattices. It has been understood for some time\cite{rotatedGraphite} that in graphite
$\theta$ can depart from Bernal values.  With some interesting exceptions\cite{ruthenium,vanHove_Andrei},
most recent studies of inter-layer twists in FLG have focused on samples grown on
SiC\cite{epitaxialGraphene}.  In particular Hass {\em et. al.} have demonstrated that orientational disorder
is normally present in carbon-face SiC epitaxial FLG samples\cite{rotatedEpitaxial}.
The present work is motivated primarily by the need to achieve a more complete understanding of transport
in these graphitic nanostructures, which currently appear to provide the
most promising platform for applications.
\begin{figure}[h]
\includegraphics[width=0.7\linewidth]{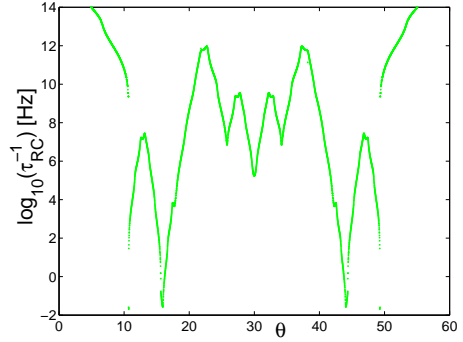}
\caption{Interlayer (RC) equilibration rate as a function of twist angle $\theta$.  These
results were calculated for two layers with equal carrier densities ($n=5 \times 10^{12} cm^{-2}$) and $\epsilon_{\ty F}\tau=3$,
where $\epsilon_{\ty F}$ is the Fermi energy and $\tau$ is the isolated-layer
Bloch state lifetime.  The relaxation rate is dominated by separate features that appear near every commensurate
angle, but differ in strength by many orders of magnitude.  The tails of individual features
have been cut-off in this plot in order to
reveal weaker features that will emerge in more ideal bilayers.  Except near $\theta=0$, the
equilibration rate is surprisingly slow for two layers separated by an atomic length scale.}
\label{fig:tauRC}
\end{figure}

In a bilayer system, the relative rotation angle $\theta$ can be classified as either commensurate or incommensurate\cite{ShallcrossLong}.
In the former case the misaligned bilayer system still forms a crystal, albeit one with
larger lattice vectors and more than four atoms per unit cell.  Commensurability occurs at a countably infinite
set of orientations; but the probability that a randomly selected orientation angle is commensurate vanishes.
The energy bands of commensurate twisted multilayers disperse approximately linearly with momentum \cite{ARPESepitaxial,Santos,Shallcross},
except at energies very close to the Dirac point. However, the Dirac velocity is reduced compared to that of a single layer
system especially for rotation angles close to $0^{\circ}$ or $60^{\circ}$\cite{Santos,vanHove_Andrei}.
The linear Dirac-like dispersion contrasts with the approximately quadratic dispersion found in a Bernal stacked bilayer system\cite{McCann}.
Incommensurate bilayers are not crystalline and therefore their
electronic properties  cannot be analyzed using Bloch's theorem.

Here we develop a theory of the vertical transport properties of twisted FLG samples
which is valid in the incoherent transport limit\cite{McKenzie}.
We show that the specific linear conductance between misaligned layers is enhanced over a small
but finite range of twist angles near those that produce relatively short period commensurate structures, that the
conductance peak angles shift with in-plane magnetic field $B_{\parallel}$, and that the
peaks become narrower and stronger when the isolated layer Bloch state lifetime $\tau$ increases.
The differential conductivity tends to be negative near commensurate conductance peaks and positive otherwise.
Typical theoretical results for the dependence of the interlayer equilibration rate on $\theta$ are presented in Fig.~\ref{fig:tauRC}.
In the following we first explain the analysis which supports these statements and then discuss some implications for FLG electronics.

Studies of transport between weakly coupled two-dimensional (2D) electron systems have a long
history\cite{2D2DTunnelTheory,2D2DTunnelExpt} in semiconductor heterojunctions systems.
In that case epitaxial tunnel barriers are responsible for nearly perfect 2D momentum conservation, which then
helps to make vertical transport a powerful probe of electronic properties.  Our theory of vertical transport in FLG
is similar to the successful semiconductor heterojunction theory\cite{2D2DTunnelTheory}.
We derive an expression for tunneling current $I$ {\em vs.} bias voltage $V$ by using a $\pi$-orbital tight-binding model,
approximating inter-layer hopping processes at leading order in perturbation theory,
and accounting for the inevitable presence of a finite disorder potential which limits the life-times of
Bloch states in each layer.  These steps lead to
\bea
I(\theta) &=& e g_s \int \frac{d\omega}{2\pi}   \left[ n_{\ty F1}(\omega) - n_{\ty F 2}(\omega+eV) \right]  \nonumber \\
&& \sum_{\bm{kp'}} |T^{{\alpha\beta}}_{\bm{k p'}}|^2 A_{1\alpha}(\bm{k},\omega) A_{2\beta}(\bm{p'},\omega+eV),
\label{I}
\eea
where $g_s=2$ accounts for spin degeneracy, $A_{i\alpha}(\bm{k},\omega)$ is the spectral function for band $\alpha$ and layer $i$,
$n_{{\ty F}i}$ is the Fermi distribution function for layer $i$,
and $T^{\alpha\beta}_{\bm{k p'}}$ is the tunneling matrix element
between isolated layer Bloch states with
band and crystal momentum labels, $|\bm{k}\alpha\rangle$ and $|\bm{p'}\beta\rangle$.
The sums over $\bm{k}$ and $\bm{p'}$ may be taken over the unrotated and rotated Brillouin zones respectively.
We derive Eq.~(\ref{I}) in section 2 of the Supplementary Information , where we justify its neglect of disorder vertex-corrections.
In our calculations, $A$ is approximated by a Lorentzian function with
full-width-half-maximum $\hbar/\tau$ centered on the band energy $\epsilon_{i\alpha}(\bm{k})$.
(Hereafter $\hbar=1$ and length is measured in units of $a_c=1.42\AA$, the carbon-carbon distance in graphene.)  Eq.~(\ref{I}) is valid in the weak tunneling regime in which
$T$ is smaller than life-time broadening $1/\tau$, allowing coherent tunneling processes
to be neglected.  This condition is satisfied in typical samples except at rotation angles very
close to $0^{\circ}$ or $60^{\circ}$.

In a twisted bilayer system the tunneling matrix element depends strongly on the relative orientation of the two graphene sheets.
The honeycomb lattice vectors of the rotated layer $\bm{R'}$ are related to those of the unrotated layer $\bm{R}$ by
$\bm{R'}=M(\theta)\bm{R} + \bm{d}$.  Here $M$ is the transformation matrix for rotations in the lattice plane and $\bm{d}$ is a translation vector.
Corresponding rotations occur in reciprocal space so that $\epsilon_{1\alpha}(\bm{p}) =\epsilon_{2\alpha}(\bm{p'})$ when $\bm{p'}=M(\theta) \bm{p}$.
Commensurability is determined only by $M$, but linear translations of one layer relative to the other do modify $T$, and hence
the tunneling current.

The magnitude of $T$ depends on the $\pi$-orbital interlayer hopping amplitudes of our tight-binding model
which we estimate using a simple two center approximation scheme explained in section 1 of the
Supplementary Information.  We find that
\bea
T^{\alpha\beta}_{\bm{kp'}} &=& \frac{1}{\Omega_0} \sum_{s,\bar{s}}(a_{\bm{k}s}^{(\alpha)})^{\star} a_{\bm{p}\bar{s}}^{(\beta)}
 \sum_{\bm{G_1}\bm{G_2}} t_{\bm{k+G_1}} e^{-i(\bm{k+G_1) \cdot d}} \nonumber \\
 &\times&   e^{i\bm{G_1}\cdot\tau_s} e^{- i\bm{G_2} \cdot \tau_{\bar{s}}}  \;  \delta_{\bm{k+G_1,p'+G_2'}}     \label{T_kSpace}
\eea
where $\Omega_0$ is the area of a unit cell.
Here $\bm{G_1}$ and $\bm{G_2}$ are summed over reciprocal lattice vectors,
primed wavevectors are rotated, $s$ and $\bar{s}$ label the two
triangular honeycomb sublattices centered at positions $\bm{\tau_s}$, and
$a_{\bm{k}s}^{(\alpha)}$ is the sublattice projection of the $|\bm{k}\alpha\rangle$
Bloch state in the unrotated layer.  In Eq.(\ref{T_kSpace}), which is derived in section 1 of the Supplementary Information,
$t_\bm{k}$ is the 2D Fourier transform of the finite-range
inter-layer hopping amplitude.  As we will explain, the interlayer conductance and
the layer equilibration rate are proportional to $|t_\bm{k}|^2$ values for $|\bm{k}|$'s that are
larger than the Brillouin-zone scale (except for $\theta\approx 0^\circ, 60^\circ$).  Because the inter-layer distance is already larger than the
carbon-carbon distance within a layer, these $|t_\bm{k}|^2$ values tend to be both extremely small and extraordinarily
sensitive to details of the inter-layer tunneling model that are otherwise inconsequential.

\begin{figure}[h]
\includegraphics[width=1\linewidth]{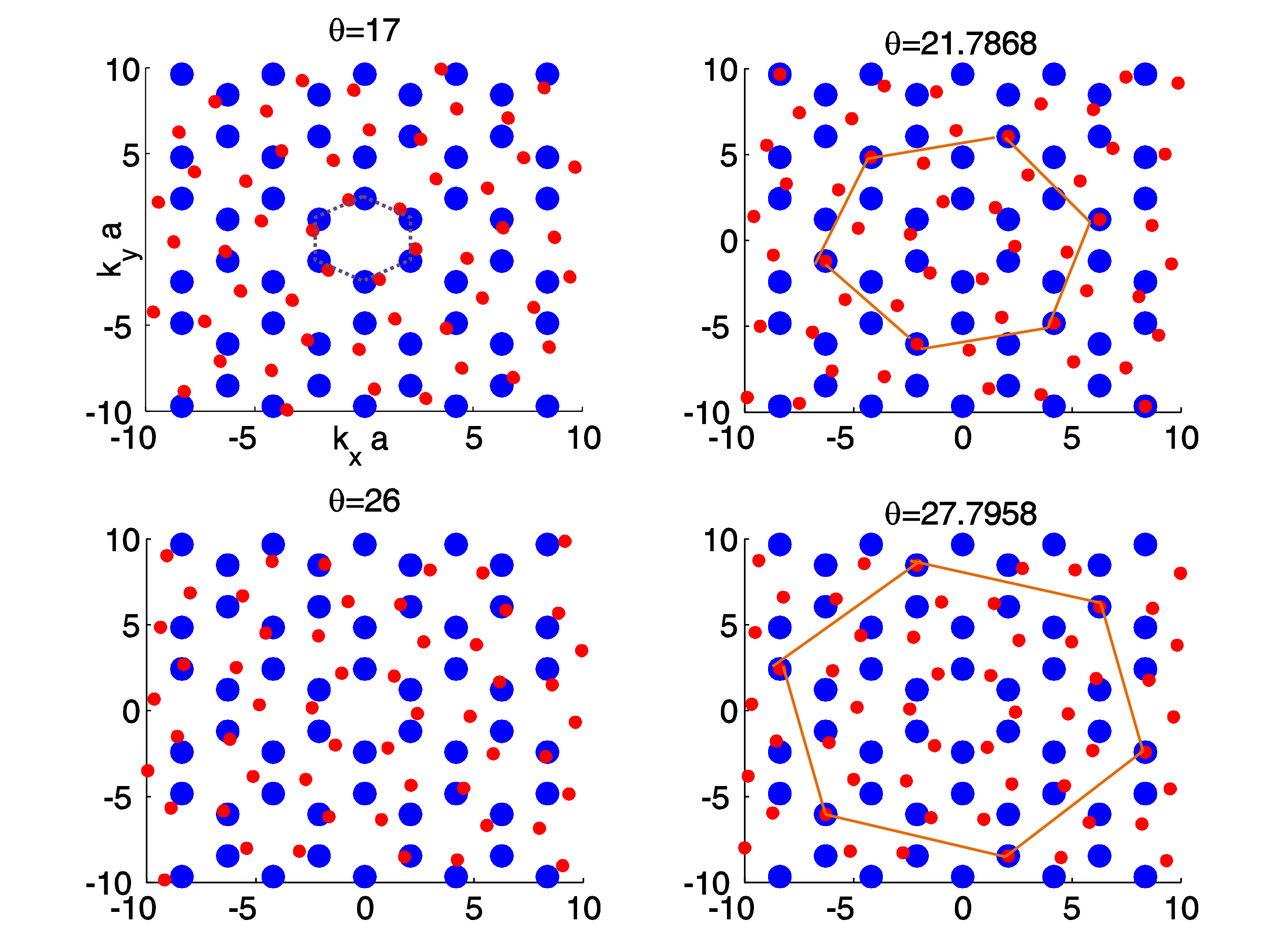}
\caption{Fermi circles in an extended zone scheme. The blue(large) and red(small) circles correspond to the
Fermi circles in the unrotated and rotated layers respectively.  The area enclosed by the circles is proportional
to the carrier density.  Conductance contributions occur when the Fermi circles intersect and are much larger when the
intersection occurs closer to the origin of momentum space.
The Brillouin-zone boundary connects the centers of the inner shell of blue circles as
indicated by the dashed lines in the $\theta=17^{\circ}$ panel.}
\label{fig:Gextended}
\end{figure}

We have used Eqs.(\ref{I}) and (\ref{T_kSpace}) to evaluate interlayer currents as a function of
rotation angle $\theta$, carrier density, bias voltage, and disorder strength.
Since for typical electronic densities the temperature $T$ is much less than the Fermi temperature
we focus on $T=0$ hereafter.
It is helpful to first focus on the linear conductance
\be
G(\theta) = \frac{e^2 g_s}{2\pi} \sum_{\bm{kp'}} |T^{{\alpha\beta}}_{\bm{k p'}}|^2 A_{1\alpha}(\bm{k},\epsilon_{\ty F}) A_{2\beta}(\bm{p'},\epsilon_{\ty F}).
\label{G}
\ee
The equilibration rate plotted in Fig.\ref{fig:tauRC} was obtained by viewing the bilayer as a
leaky capacitor and ignoring any screening by graphene $\sigma$ orbitals.  This model yields an
RC circuit with time constant $\tau_{\ty {RC}}$ related to the conductance by
$G/\cA =0.027 \; \tau_{\ty{RC}}^{-1}$ where $\cA$ is the layer area in $m^2$, $G$ is measured in Siemens, and $\tau_{\ty{RC}}$ in seconds.
Apart from a change in scale, Fig.\ref{fig:tauRC} can then be viewed as a plot of the interlayer conductance.
We find that the tunneling conductance increases abruptly near commensurate angles,
that the height of the peaks scales linearly with $\epsilon_{\ty F} \tau$ (for $\epsilon_{\ty F}\tau>1$), and
that the peaks narrow as $\tau$ increases. The discontinuous jumps of $\log(G)$ in Fig.\ref{fig:tauRC} are artificial
and result from a numerical procedure in which momenta $\bm{k}$ and $\bm{p'}$ in Eq.(\ref{G}) are restricted to the vicinity of the Fermi energy.
This procedure suppresses the tails of all commensurate features, allowing more minor features to be revealed.
In practice the conduction tails corresponding to highly commensurate structures will dominate $G$ over a range of angles
that depends on $\tau$.
Limited by computational power we considered $\tau^{-1} \approx 75 \ meV$ in Fig.\ref{fig:tauRC} however in
epitaxial graphene the lifetime can be more than an order of magnitude longer\cite{stroscio}.
An accurate theory of the conduction-peak tails would require a reliable theory of the isolated-layer spectral function tails.

Why is the tunneling conductance enhanced at commensurate rotation angles?
To understand the relation between interlayer current and commensurability it is illuminating to
plot the Fermi surfaces of both layers, periodically extended in momentum space by
adding reciprocal lattice vectors to the crystal momenta of the electrons.
As we see in Eq.(~\ref{T_kSpace}), allowed interlayer tunneling processes are diagonal in this generalized momentum.
The left panels in Fig.\ref{fig:Gextended} corresponds to the incommensurate rotation angles $\theta=17^\circ,26^\circ$ whereas the right
panels correspond to the commensurate angles near $\theta=21.8^\circ,27.8^\circ$. We use different Fermi surfaces sizes for clarity; similar
considerations apply independent of the sign or magnitude of the carrier density ratio.
The key feature to notice in these plots is that at commensurate rotation angles some Fermi spheres overlap.
Overlaps of circles centered on the extended Dirac points,
{\em always} accompany commensurate real-space structures because the set of extended Dirac points
forms a momentum space honeycomb lattice that differs from the real space honeycomb lattice
only by a scale factor and by a rotation.
If overlaps occur in real-space, they also occur in momentum space.  Notice that this property holds only
when the Brillouin-zone corners are extended to fill momentum space; if the Dirac point occurred elsewhere in
the isolated layer Brillouin-zone, the dependence of inter-layer conductance on $\theta$ would be quite different.

\begin{figure}[h]
\includegraphics[width=0.8\linewidth]{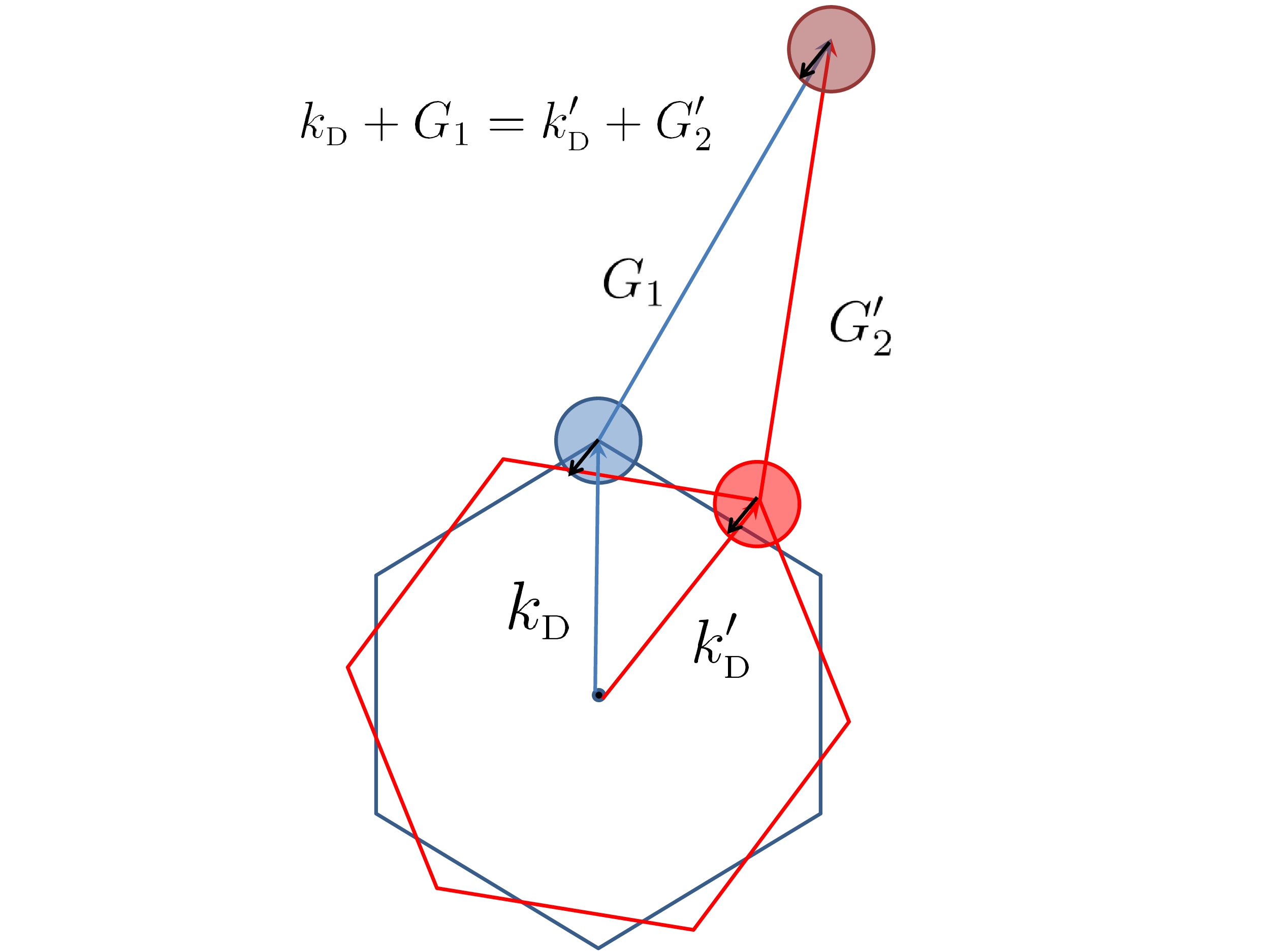}
\caption{Nesting of Dirac cones at commensurability. For commensurate rotation angles every momenta state on the rotated Fermi circle is mapped onto a momenta
state of an unrotated Fermi circle.}
\label{fig:nesting}
\end{figure}

The overlap of extended Dirac points does not fully explain the conductance peaks at finite
density, since Fermi energy states at finite carrier density are displaced from the Dirac point.
The nesting between Fermi surfaces alluded to in Fig.\ref{fig:Gextended} actually depends not only on commensurability, but also on the
fact that for typical carrier densities the Fermi surface is well approximated by a circle centered on the
Brillouin-zone corners. For equal densities then, matching Dirac points implies complete Fermi surface nesting (see Fig.\ref{fig:nesting}).
When the two-layers have different densities, the peak conductance will not occur at the nesting angle;
instead the conductance will have a double-peak structure with features offset to both sides of the
commensurate angle.

Commensurate rotation angles can be classified as either inter-valley or intra-valley. In the former the two
Dirac points $k_{\ty D}$ and $k'_{\ty D}$ that coincide in the extended momentum picture are associated with different valleys (in the aligned bilayer) whereas for
intra-valley rotation angles they belong to the same valley. An inter-valley commensurate rotation
is illustrated in  Fig.\ref{fig:nesting}.

Away from commensurate angles the energy difference between states which have the
same extended momentum is typically much larger than the Fermi energy,
and the spectral function width $1/\tau$
(see left panels of Fig.\ref{fig:Gextended}).
The conductance is therefore very small away from the commensurate-angle peaks.
The Dirac-like linear spectrum of an ideal commensurately twisted bilayer does not,
as is commonly stated, indicate that the ideal twisted layers are decoupled.
At commensurate angles the perfect crystal wavefunctions near the Dirac point
are in fact coherent equal weight contributions from the two layers.
In the limit of large in-plane Bloch state lifetimes, the conductance becomes very large
and eventually the incoherent transport picture will fail.

As we have explained, vertical transport at commensurability is dominated by processes in which
an electron tunnels from a momentum near a Dirac point of one layer, to a momentum that is the
same distance from a Dirac point of the other layer.
Since carrier densities per unit cell are always small,
we can replace $t_{\bm{k+G}}$ in Eq.(~\ref{T_kSpace}) by $t_{\bm{k_{D}+G}}$
where $\bm{k_{\ty D}}$ is the Dirac point momentum.
We then find that the conductance peak can be expressed as the product of
geometry-related and phase space factors:
\be
G \approx R(\theta_c,\bm{d}) \sum_{\bm{k}} A_1(\bm{k},\epsilon_{\ty F}) A_1(\bm{k},\epsilon_{\ty F}),     \label{GRAA}
\ee
where $\theta_c$ is the commensurate orientation. $R(\theta_c,\bm{d})$ depends mainly on the value of
$t_{|\bm{k_{D}+G}|}$ at which the extended Dirac points overlap (see Fig.\ref{fig:Gextended}), while the remaining phase space
factor is identical to the one that appears in the theory of coupled quantum wells\cite{2D2DTunnelTheory}.
For equal densities in the two layers, the Fermi surfaces nest precisely. For pure rotations $R$ can be calculated analytically.
For inter-valley commensurate rotation angles we find that
\be
G(\theta_c) = \cA g_s g_v \frac{e^2}{\hbar} \frac{\epsilon_{\ty F}\tau E_g^2(\theta_c)}{16 \pi v^2}.     \label{GD}
\ee
Here $E_g$ is the energy difference between the top conduction band and bottom valence band of the
twisted bilayer at the Dirac point, and we assumed that $\epsilon_{\ty F}\tau>1$.
In section 4 of the Supplementary Information we derive Eq.(\ref{GD}) and obtain a similar formula for intra-valley rotation angles.
In addition we numerically verify that the conductance changes only by a factor of order unity as $\bm{d}$ is varied across the unit cell.
Eq.(\ref{GD}) therefore provides a good estimate for $G$ regardless of the relative translation between the two layers (see section 3 of the Supplementary
Information).

When the densities differ, Fermi circles in different layers begin to overlap near $\theta=\theta_c$ only after a momentum-space relative shift
$\bm{Q}$ equal in magnitude to the difference of the two Fermi wavevectors.  As in
semiconductor double-wells\cite{2D2DTunnelTheory,2D2DTunnelExpt,Lyo}, a shift $\bm{Q}=\bm{\hat{z}}\times\bm{\hat{e}}d_\perp/l^2_{\ty H}$, where $l_{\ty H}$ is the
magnetic length, can be accomplished in a bilayer with layer separation $d_\perp$ by applying an in-plane magnetic field $B_{\parallel}\bm{\hat{e}}$.  For graphene bilayers, however, a relative
momentum space shift can also be achieved by rotation, as is clear from Fig.\ref{fig:Gextended}.  For small
departures from commensurability $\bm{Q} \approx (\theta-\theta_c) \; \hat{z} \times
(\bm{k_{D}+G})$.  For equal densities, both rotations and in-plane fields dramatically suppress
the conductance peak
when $v Q \geq 1/\tau$ where $v$ is the band velocity of graphene.
For example, for $n=4 \cdot 10^{12} cm^{-2}$ and $\tau=50$ fsec \cite{scattringTime}, the conductance
peak should nearly completely disappear at $0.15$ Tesla.
FLG should therefore provide a palette on which gate voltages, in-plane magnetic fields, and
rotations can be mixed to produce a rainbow of interrelated and extraordinarily strong magnetic-field and
strain sensitive resistance effects.

\begin{figure}[h]
\includegraphics[width=0.5\linewidth]{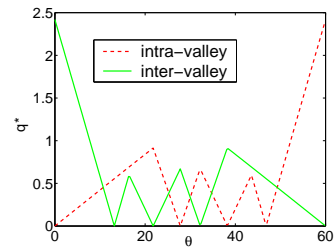}
\caption{The minimum separation between extended Dirac points $q^\star$ as a function of rotation angle $\theta$.}
\label{fig:qStar}
\end{figure}
In general the interlayer conductance $G(\theta)$ is peaked whenever any extended Fermi surface overlap occurs at
reasonably small reciprocal lattice vectors.  The degree of overlap
can be parameterized by $q^\star$, the minimum separation between extended Dirac points of the rotated and unrotated layers
for a reciprocal lattice vector truncation chosen to reflect the scale on which $t_{\bm{q}}$ falls off.
For a clean system, tunneling conductance at equal density is appreciable as long as  $q^\star
\approx |\theta-\theta_c| |\bm{k+G}| < 2k_{\ty F}$.  Because $2k_F$ in FLG electronic systems is always small compared
to reciprocal lattice vector scales, the conductance peaks are invariably sharp when plotted as a function of $\theta$.
As an example $q^\star \approx 6.39 |\theta-\theta_c|$ in the vicinity of $\theta_c=27.8^\circ$ for the reciprocal
lattice vector illustrated in Fig.\ref{fig:Gextended}.
In Fig.\ref{fig:qStar}, $q^\star$, minimized over the first two G-shells, is plotted
as a function of angle.
Overlap between the Fermi spheres of the two layers
will therefore persist over the angle range for which  $q^\star$ is smaller than $2k_F$.

Electronic structure calculations for ideal commensurate bilayers
demonstrate that $E_g$ decreases very rapidly as the number of atoms per unit cell increases\cite{ShallcrossLong}. $E_g=780$ meV
for a unit cell of 4 atoms, and already less that 1 meV for a unit cell of 100 atoms.
It is therefore plausible that conductance tails that correspond to the few most lowest order commensurate
angles (e.g. $\theta_c=0^\circ,21.8^\circ,27.8^\circ,32.2^\circ,38.2^\circ,60^\circ$) will dominate $G$
at every rotation angle.  Eqs.(\ref{GRAA}) and (\ref{GD})
should therefore be interpreted as a lower bound for the conductance at higher order commensurate $\theta_c$'s.

We now turn to the non-linear $I-V$ of twisted bilayer graphene. At zero temperature
\be
I(\theta,V) = \frac{e g_s}{2\pi} \sum_{kp} |T_{kp'}|^2 \int_{\epsilon_F-eV}^{\epsilon_F} d\omega A_1(k,\omega) A_2(p',\omega+eV).  \label{IkpT0}
\ee
We numerically find that the I-V curves at commensurate
and incommensurate angles are drastically different. At relatively small bias
voltage the currents corresponding to commensurate angles are several orders of magnitude larger than their incommensurate counterparts.
On the other hand negative differential conductances invariably appear at commensurate angles, whereas $dI/dV$ tends
to be small and positive at incommensurate angles.

In classic tunneling experiments, a bias voltage induces an equal electric potential difference between the layers.
Total energy conservation then implies kinetic energy changes equal to $eV$ upon tunneling.
Since, as we have explained, the allowed tunneling processes at commensurate angles are between states with the
same kinetic energy bias voltages tend to decrease tunneling currents.
Following the same approximations that led to Eq.(\ref{GRAA}) we can capture this effect mathematically
by expressing the interlayer current in product form:
\bea
I&=&\frac{e}{2\pi}\sum_{\alpha\beta}R^{\alpha\beta}(\theta,\bm{d}) \int_{-\infty}^{\infty} d\omega \left[ n_{\ty F}(\omega)-n_{\ty F}(\omega+V) \right]  \nonumber \\
&\times& \sum_{\bm{K}} A_{\alpha}(k,\omega) A_\beta(k,\omega+eV)  \label{IV_commensurate}
\eea
In Eq.(~\ref{IV_commensurate}) we have allowed for both intraband and interband tunneling at large biases.
As long as $eV < \epsilon_{\ty F}$
tunneling between conduction bands dominate $I$ when both layers are n-type.
In this intermediate non-linear regime
the two Lorentzian shaped spectral functions in Eq.(\ref{IV_commensurate}) overlap only weakly and
\be
I(\theta_c,\bm{d}) \approx G(\theta_c,\bm{d}) \frac{V}{1+(eV\tau)^2}.       \label{IsmallV}
\ee
for $\epsilon_{\ty F} \tau > 1$. Negative differential conductance occurs when $eV\tau>1$.
For incommensurate twist angles, crystal momenta conservation can not be sustained at the Fermi surface. Increasing $V$ unblocks
processes in which tunneling occurs between states with different kinetic energies and leads to a slow increase of the tunneling current
with a complex dependence on $t_{\bm{q}}$ and $\hbar/\tau$.
For $eV > \epsilon_{\ty F}$, the current increases monotonically with $V$ for both commensurate and incommensurate twist angles.
The commensurate tunneling current has a sharp rise at $eV=2\epsilon_{\ty F}$ due to momentum conserving processes allowed at high bias voltage in which a valence band electron
in one layer tunnels to the conduction band of the opposite layer. For commensurate angles it follows from (\ref{IV_commensurate}) that these inter-band processes
eventually dominate the tunneling current and that
\be
I \approx \frac{e^2}{4v^2} R^{vc} \Theta(V-2\epsilon_{\ty F}) V     \label{IlargeV}
\ee
to leading order in $1/V \tau $. Here $\Theta$ is the Heaviside step function. The finite temperature
corrections to Eqs.(\ref{IsmallV},\ref{IlargeV}) are exponentially small in $T/\epsilon_{\ty F}$.

The extension of our theory to FLG is straightforward in the linear regime.
In the simplest case each layer is rotated with respect to its neighbors sufficiently to drive the
system into an incoherent transport regime.  The weak links between layers then act like classical resistors
which appear in series in vertical transport.  The resistance of each link depends on the rotation angle between layers
and on the densities in both layers.  We anticipate a very rich and complex behavior of FLG in the non-linear regime.
The negative differential conductivities are likely to give rise
to steady state multistability and to chaotic temporal response,
as occurs in semiconductor multiple-quantum-well systems\cite{SCsuperlattices}.
A more complicated scenario could arise in turbostratic graphene. There
the entire layered structure is composed of a set of coherent multi-layer substructures, characterized by either a Bernal or an AA stacking sequence.
Weak links which play a dominant role in limiting vertical conductance appear due to occasional twists.
The calculations for the resistance of each twisted interface
closely follow those outlined above for the two-layer case when supplemented by a band index
for the various 2D energy bands of a coherent substructure.

One application of our theory is to assess whether or not twisted graphene layers are effectively isolated from an electrical point of view.
The equilibration time between layers that are spatially uniform but out of equilibrium is plotted in Fig.~\ref{fig:tauRC} and
is very long compared to characteristic electronic time scales for rotation angles far from important commensurabilities,
near $10^\circ$ for example.  The steady-state equilibration length between separately contacted layers can be estimated by
equating inter-layer conductances, which are proportional to sample area, with the intra-layer conductance per square. For the commensurate angle $\theta_c=21.8^\circ$,
for example, the sample area at which they are identical is approximately
$0.04 \mu m^2$. As evident from Fig.\ref{fig:tauRC}, the corresponding areas for small rotation angles near the
AA and AB stacking sequences are even smaller. For small rotation angles, the two layers are therefore strongly coupled.

Finally we remark that the extraordinary sensitivity of the tunneling conductance
to the twist angle found here suggests that misaligned graphene bilayers might be useful as ultra-sensitive strain gauges or pressure sensors\cite{bunch}
which are widely used in biological, mechanical  and optical systems.

The authors acknowledge support from CERA, SWAN and the Welch Foundation and helpful conversations with  W. de Heer, R. Duine, P. First,
D. Goldhaber-Gordon, R. Lifshitz, and E. Tutuc.

\begin{widetext}

\vspace{5mm}

\centerline{\Large \textbf{Supplementary Information}}

\section{The tunneling matrix elements}

The interlayer hopping terms in a $\pi$-band tight-binding Hamiltonian for twisted graphene bilayers depend in
general on the positions of all carbon atoms.
Our analysis of inter-layer conductance and equilibration is based on a simple two-center model in which the interlayer hopping parameter between two sites, $t(\bm{r})$,
depends only on the planar projection of their separation $\bm{r}$.  In the main text we used an equation, derived below, which relates
the inter-layer hopping amplitudes of twisted bilayers to $t_\bm{q}$, the two-dimensional Fourier transform of $t(\bm{r})$.

One strategy which can be used to estimate $t(\bm{r})$ is to assume functional forms for the
distance dependence of the Slater-Koster $t_{pp\sigma}$ and $t_{pp\pi}$ hopping functions\cite{SlaterKoster},
and then fit them to accurately known parameters of untwisted bilayers.  We have explored this
approach, following the procedures adopted in Refs. \cite{Pereira,ShallcrossLong}, but
have concluded that it tends to underestimate hopping amplitudes near the Dirac points of twisted bilayers.
We have therefore decided to obtain numerical estimates by directly fitting an
{\em ansatz} for $t_\bm{q}$ to obtain
\be
t_\bm{q} = t_0 \; e^{-\alpha (q d_\perp)^\gamma} ,     \label{tq}
\ee
where $t_0=2 \ eV \AA^2$, $\alpha=0.13$, $\gamma=1.25$, and $d_\perp=3.34 \AA$ is the distance between the layers.  The value used for $t_0$
is the average of values implied by the models in Refs.\cite{Pereira,ShallcrossLong}.
Since $t_0$ is the sum of all inter-layer hopping parameters, it should be estimated reliably by any
parameterization that uses accurate values for the largest hopping parameters.
We fix $\alpha$ and $\gamma$ so that the values of the ideal bilayer gaps are accurate for the lowest order
commensurate structures.  These are proportional to $t_{k_D}$ ($\theta=0^\circ$ and $\theta=60^\circ$)
and $t_{6.4/a_c}$ ($\theta= 21.8^\circ$ and $\theta=38.2^\circ$) where $a_c=1.42 \AA$ is the carbon-carbon distance in single layer graphene.
See details in Sec.~\ref{sec_Gcommensurate} below. We fit the energy gaps to values extracted from the {\em ab initio}
calculations by Shallcross {\em et al.}\cite{ShallcrossLong}.  Note that these values of $t_\bm{q}$ characterize
short-distance roughness in the inter-layer hopping landscape which survives Fourier transformation at
large wavevectors, which is not simply related to typical inter-layer hopping strengths.
The energy gaps that we obtain at $\theta=21.8^\circ$ using the real space parameterizations of
$t_{pp\sigma}(r)$ and $t_{pp\pi}(r)$ in Refs. \cite{Pereira,ShallcrossLong} are both substantially smaller than
the {\em ab initio} gaps of Shallcross {\em et al.}\cite{ShallcrossLong}.

We now derive the expression for the hopping amplitude between Bloch states in twisted
bilayers that is used in the main text.
The Bloch state in layer $j$ with crystal momentum $\bm{k}$ and band index $\alpha$ can be written as
\be
|\Psi^{(j)}_{\bm{k}\alpha}\rangle = a^{j\alpha}_{\bm{k}{\ty A}} \; |\psi_{\bm{k}{\ty A}}^{(j)}\rangle + a^{j\alpha}_{\bm{k}{\ty B}} \; | \psi_{\bm{k}{\ty B}}^{(j)}\rangle
       \label{Bloch_wf}
\ee
where $A$ and $B$ label the two triangular honeycomb sublattices,
\be
\left(
  \begin{array}{c}
    a^{1\alpha}_{\bm{k}{\ty A}} \\[0.1 cm]
    a^{1\alpha}_{\bm{k}{\ty B}} \\
  \end{array}
\right) =
\frac{1}{\sqrt{2}}
\left(
  \begin{array}{c}
    e^{i\Theta_\bm{k}} \\[0.1 cm]
    \alpha \\
  \end{array}
\right),
\ee
and $\Theta_\bm{k}$ is the phase of the inter-sublattice hopping term in the single-layer tight-binding model.
For nearest neighbor hopping within the planes $\Theta_\bm{k} = {\rm arg}\lp\sum_j e^{i \bm{k} \cdot \bm{\delta}_j}\rp$
where the $\bm{\delta}_j$ are the three vectors
connecting an atom with its nearest neighbors. The Bloch state projection on
sublattice $s$ is
\be
| \psi^{(1)}_{\bm{k}s}\rangle = \frac{1}{\sqrt{N}}\sum_{\bm{R}} e^{i\bm{k} (\bm{R}+\bm{\tau}_s)} |\bm{\tau}_s+\bm{R}\rangle,         \label{wf1_sublattice}
\ee
where $|\bm{\tau}_s+\bm{R}\rangle$ is a site-representation basis function of the tight-binding model.
In Eq.(\ref{wf1_sublattice}) $\bm{R}$ is a triangular lattice vector,
$N$ is the number of unit cells in the system, and we choose $\tau_{\ty A}=0$ and $\tau_{\ty B}$
equal to the vector connecting the two atoms within a unit cell.

The relative orientation of the two layers can be described by a rotation matrix $M(\theta)$ and a translation vector $\bm{d}$.
Therefore every Bloch wave function in the second layer
is related to a Bloch wave function in the first layer by
\be
| \Psi_{\bm{k}'\alpha}^{(2)}\rangle = | \Psi_{\bm{k}\alpha}^{(1)} \rangle        \label{psi122}
\ee
with $|\bm{R}+\bm{\tau}_s\rangle$ in layer 1 replaced by $|\bm{R'}+\bm{\tau}_s'\rangle$ in layer 2,
$\bm{r'}=M \bm{r + d}$ for all positions and $\bm{k'}=M \bm{k}$.
Using primes to indicate layer 2 variables and invoking the two-center approximation for the
inter-layer tunneling amplitude,
\be
\langle \bm{\tau}_s+\bm{R}| H_{inter} |\bm{\tau'}_s+\bm{R'}\rangle \; = \; t( \bm{\tau}_s+\bm{R} - \bm{\tau'}_s- \bm{R'}),
\ee
we find that
\be
\langle \Psi_{\bm{k}\alpha}| H_{inter}|\Psi_{\bm{p'}\beta} \rangle \equiv T^{\alpha\beta}_{\bm{kp'}} =
 \frac{1}{N} \sum_{ss'} \lp a_{\bm{k}s}^{(\alpha)} \rp^\star a_{\bm{p}s'}^{(\beta)} \sum_{\bm{R}_1 \bm{R}_2} e^{-i\bm{k} \cdot (\bm{R}_1+\tau_s)+i\bm{p} \cdot (\bm{R}'_2-\bm{d}+\tau'_{s'})}  \nonumber \\
\; t(\bm{R}_1+\tau_s-\bm{R'}_2-\tau'_{s'}).  \label{T_realSpace}
\ee
Expression (2) in the main text is obtained by Fourier expanding $t(\bm{r})$ and summing over the lattice vectors.

\section{Vertex corrections}

The general expression for the tunneling current
\bea
I(\theta,V) &=& -4e g_s \int \frac{d\omega}{2\pi} \sum T^{{\alpha\beta}}_{\bm{k_0 p'_0}} T^{\gamma\delta\star}_{\bm{k_N p'_N}}   \left[ n_2(\omega+eV) - n_1(\omega) \right]  \nonumber \\
&\times& \overline{ Im G^{\ty R}_{1\gamma\alpha}(\bm{k_N},\bm{k_0},\omega) Im G^{\ty R}_{2\beta\delta}(\bm{p'_0},\bm{p'_N},\omega+eV) }
\label{Idisorder}
\eea
is obtained using second order perturbation theory\cite{Mahan}. In Eq.(\ref{Idisorder}) $n_j$ is
the Fermi distribution in layer $j$, $G^{\ty R}_{j\gamma\alpha}$ is the retarded Green
function in layer $j$ that correspond to the propagation of a charge carrier from band $\alpha$ to band $\gamma$,
the rotation angle is $\theta$, and $V$ is the bias voltage. The over-line denotes disorder averaging. As in the main text, primed
variables are associated with the rotated layer. Since disorder breaks translation invariance, the Green functions
are not diagonal is the momentum representation.  When the disorder averages can be performed independently for the
two-layers, translational invariance is recovered and Eq.~(\ref{Idisorder}) reduces to Eq.(1) of the main text.

We average over disorder using the self-consistent Born approximation in which correlations between
the layers appear as a vertex-correction ladder diagram sum (see Fig.\ref{fig:bubble}).
For simplicity we assume white noise disorder and characterize the correlation between the disorder potentials in the two layers by $\gamma = n_i\la U_1 U_2 \ra$ where $n_i$
is the concentration of impurities and $U_j$ is the disorder potential in layer $j$. For aligned bilayers with short range tunneling we find that
\be
G = \frac{e^2 t^2 \nu_{\ty F} \tau}{2} \frac{1}{1-\gamma/\beta}        \label{Gdisorder}
\ee
where $\beta=n_i \la U_j^2 \ra$.
As evident from Eq.(\ref{Gdisorder}) the tunneling conductance diverges if the disorder potentials of the two layers are perfectly correlated.
These strong correlations are likely in a graphene bilayer because of the small distance between the layers.
The divergence of $G$ indicates the breakdown of perturbation theory, {\em i.e.} it invalidates the incoherent theory we use in this work.
A similar scenario arises for tunneling between coupled semiconductor quantum wells\cite{Zheng} when their disorder
potentials are strongly correlated.

We now show that vertex corrections are important only at very small values of the rotation angle $\theta$.
The physical origin of this behavior is twofold. First, the relevant correlation in the twisted case is between the disorder
potential in one layer and a spatially rotated counterpart in the other layer.  For any finite range disorder correlation
length, these two disorder potentials are independent making $\gamma$ in Eq.(\ref{Gdisorder}) considerably smaller.
Second, the divergence in the conductance appears due to tunneling between identical states. However, for incommensurate angles the wave vectors
of the initial and final states in a tunneling process substantially differ making $\beta$ in Eq.(\ref{Gdisorder}) considerably larger.
In the following paragraphs we explain how this latter behavior is captured in a diagrammatic perturbation theory description of a disordered system.

We first focus on the tunneling conductance for aligned layers ($\theta=0$). At zero temperature
\bea
G(\theta) &=& \frac{2e^2 g_s}{\pi} \sum T^{{\alpha\beta}}_{\bm{k_0 p'_0}} T^{\gamma\delta\star}_{\bm{k_N p'_N}}
\overline{ Im G^{\ty R}_{1\gamma\alpha}(\bm{k_N},\bm{k_0},\epsilon_{\ty F}) Im G^{\ty R}_{2\beta\delta}(\bm{p'_0},\bm{p'_N},\epsilon_{\ty F}) }.
\label{Gdisorder2}
\eea
\begin{figure}[h]
\includegraphics[width=0.3\linewidth]{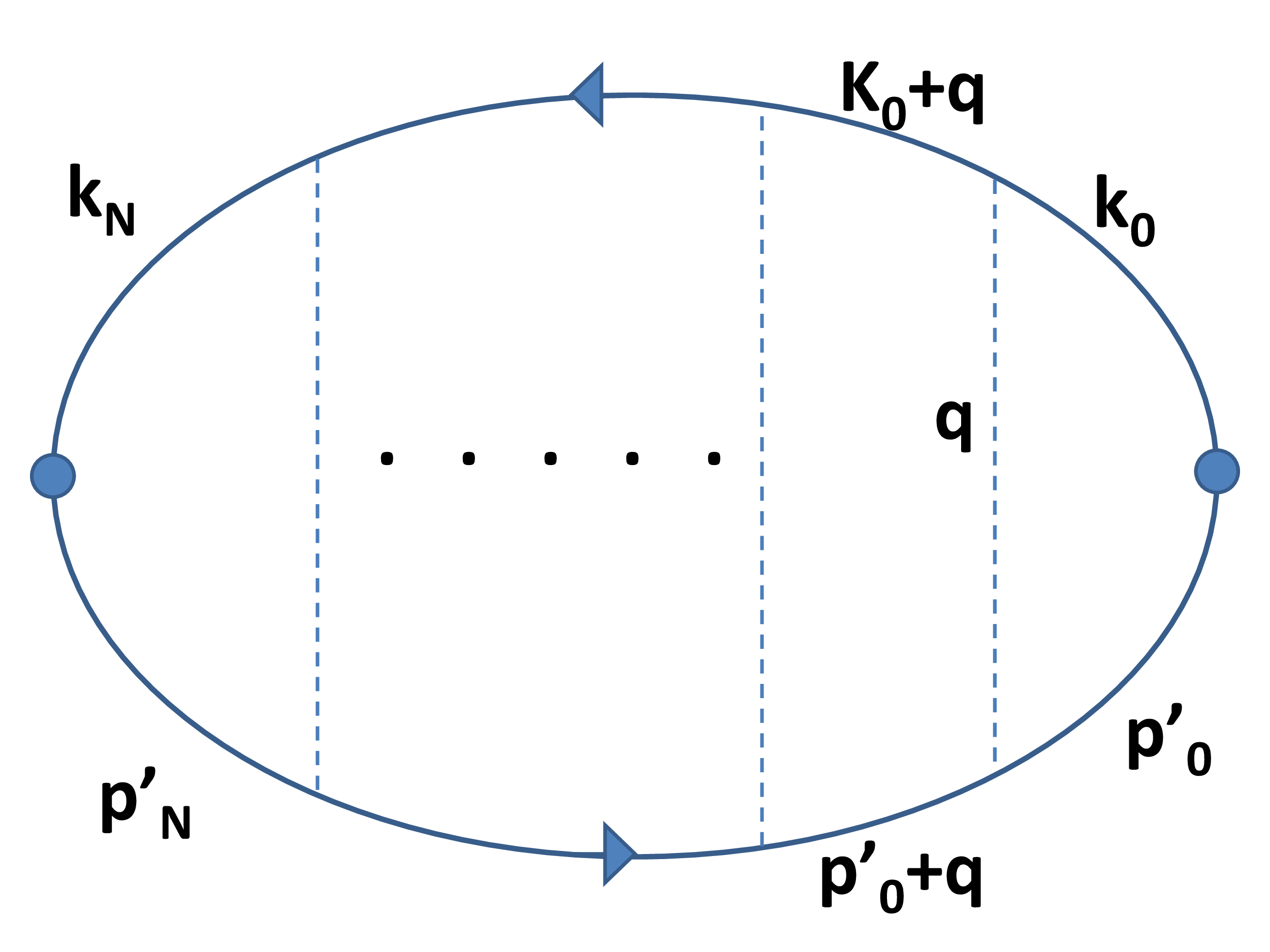}
\caption{Self consistent Born approximation. A bubble diagram with ladders.}
\label{fig:bubble}
\end{figure}
The conservation of crystal momentum in expression (2) for $T^{\alpha\beta}_{\bm{kp'}}$ implies that $\bm{p_0=k_0}$ and
that $\bm{p_N=k_N}$. For $\epsilon_{\ty F} \tau > 1$ interband transitions are inhibited so that $\alpha=\gamma$ and $\beta=\delta$.
Due to the spinor form of the wave functions each disorder line contributes $[1+\cos(\theta_{k_{j+1}}-\theta_{k_j})] \gamma/2$ to the ladder diagram.
To evaluate $\Pi^{(n)}$, the ladder diagram with $n$ disorder lines, we first integrate over the angular variables using
\be
\int_0^{2\pi} \frac{d\theta_q}{2\pi} \cos(\theta_{k_1} - \theta_{q}) \cos(\theta_{k_2} - \theta_{q}) = \frac{1}{2} \cos(\theta_{k_1} - \theta_{k_2}).
\ee
Then using $\cF(0)= 2\pi \nu_{\ty F} \tau$ where
\be
\cF(\bm{Q}) = \sum_q  G_{1\alpha}^R(q,\omega) G_{1\alpha}^A(\bm{q+Q},\omega)        \label{GRGA}
\ee
we integrate over the radial direction. In obtaining $\cF(0)$ we have replaced the energy dependent density of states by $\nu_{\ty F}$, its value at the Fermi energy.
We find that for $n \geq 1$
\be
\Pi^{(n)} =   G^\mu_1(k_0) G^\nu_2(k_0) \left[ 1 + \frac{1}{2^{n-1}}\cos(\theta_{k_0}-\theta_{k_N}) \right] \lp \frac{\gamma}{\beta} \rp^{n-1}
\frac{\gamma}{2} G^\mu_1(k_N) G^\nu_2(k_N)
\ee
where $\beta = 1/\pi\nu_{\ty F} \tau$ and $\mu,\nu=R,A$.
For $n \geq 2$ the Green functions in one layer are retarded and those of the other layer are advanced.
We now sum $\Pi^{(n)}$ to infinite order in $n$. While the sum can clearly be carried for a general tunneling matrix element $T_{kk}$
the basic physical idea is more transparent for short range tunneling. Therefore in the calculations below we assume $T_{kk}=t$ is momentum independent in which case
we recover Eq.(\ref{Gdisorder}).

We now address the role played by vertex corrections is twisted bilayers. As in the main text our discussion excludes the vicinity of $\theta=0^\circ,60^\circ$ for which $t>1/\tau$.
The procedure outlined above for calculating $G$ can be repeated for any rotation angle $\theta$. For a rotated bilayer it follows from Eq.(2) in the main text that
$\bm{k_0-p'_{0}}=\bm{G'_2-G_1} \equiv \bm{Q}$ where $Q  \approx |\bm{k_{\ty D} + G_1}||\theta-\theta_c|$.
Expression (\ref{Gdisorder}) can then be used for a rotated bilayer as well if $\beta$ is replaced by
\be
\beta_{\ty Q} = \beta F(0)/F(Q).
\ee
Because $\cF$ is a monotonically decreasing function of its argument and because $Q$ is comparable in size to the Dirac momentum  $\beta_{\ty Q} \gg \beta$.

\begin{figure}[h]
\includegraphics[width=0.5\linewidth]{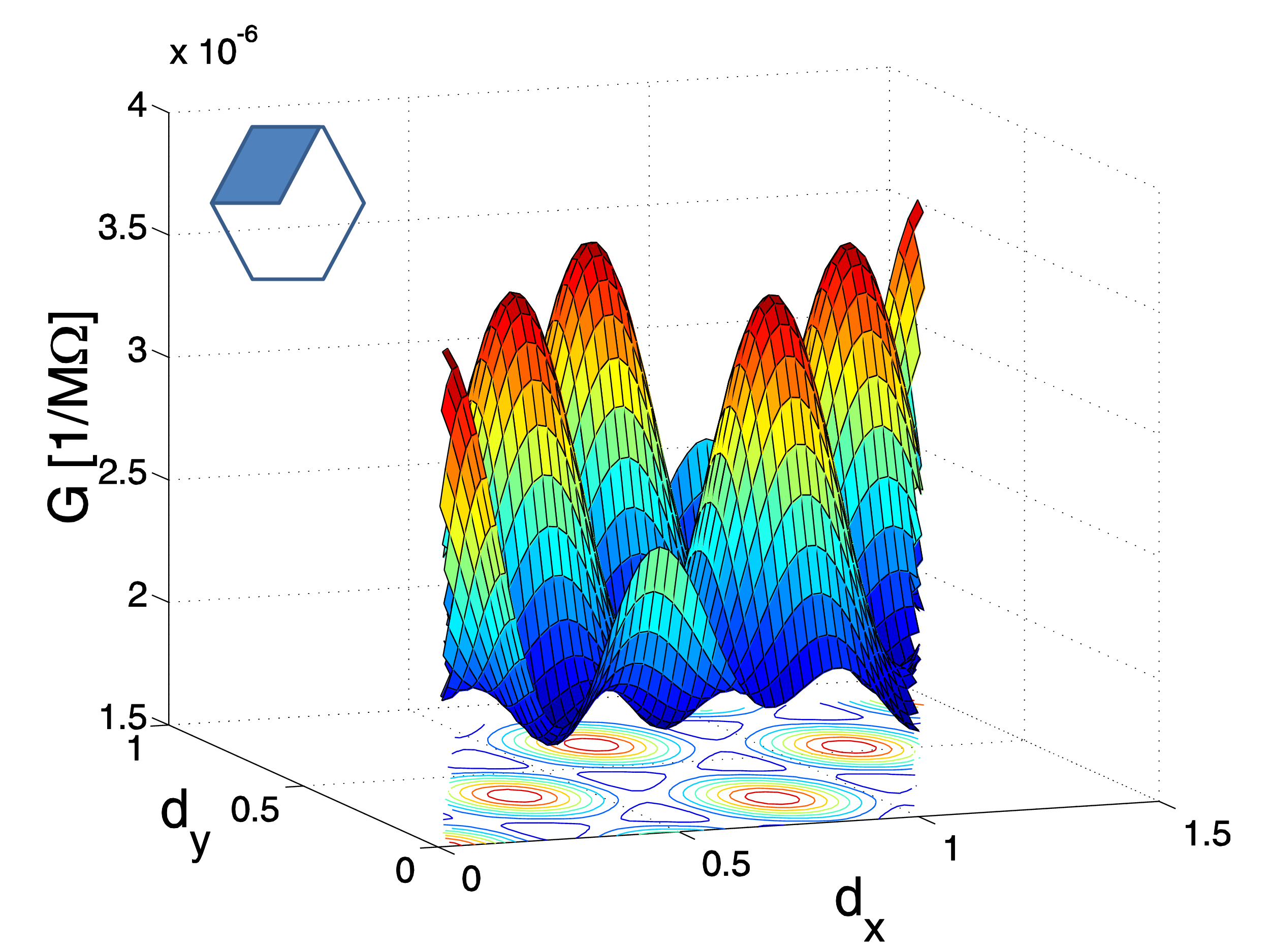}
\caption{Dependence of conductance per unit cell $G$ on translation ${\bm d}$
for $\theta=27.8^\circ$, $\epsilon_{\ty F}\tau=3$, and $n=5\cdot 10^{12} \ cm^{-2}$.}
\label{fig:Gd}
\end{figure}

\section{Dependence of tunneling current on translation}

Commensurability depends only on the relative rotation of the two graphene layers. Nevertheless linear translation of one layer with respect to the other will
change the tunneling current. In Fig.\ref{fig:Gd} the conductance at $\theta=27.8^\circ$ is plotted as a function of $\bm{d}$ for a bilayer with $n=5 \cdot 10^{12}\ cm^{-2}$ in each layer
and $\epsilon_{\ty F}\tau=3$. The dependence of the tunneling current
on $\bm{d}$ is captured by the phase factor $\exp[-i(\bm{k+G_1})\cdot\bm{d}] \simeq
\exp[-i(\bm{k_D+G_1})\cdot\bm{d}]$ in the tunneling matrix element $T_{\bm{kp'}}$.
When summed over $\bm{k}$ the result is a rapid spatial variation on the lattice constant scale, illustrated in Fig.\ref{fig:Gd}
due to the $\exp[-i(\bm{k_D+G_1})\cdot\bm{d}]$ factor, modulated by a slower variation on the Fermi wavelength scale.

As illustrated in Fig.1 the conductance peaks appear symmetrically around $\theta=30^\circ$,
but the height of a peak with $\theta < 30^\circ$ does not necessarily equal the height of the
corresponding peak at $\theta'=60^\circ-\theta$.  In fact, the relative height of the two peaks depends on $\bm{d}$.
An AA stacking sequence can be transformed to Bernal stacking either by a pure rotation with $\theta=60^\circ$ or by a translation
with $\bm{d}=(1,0)$. Since the latter transformation does not influence commensurability
any commensurate angle $\theta_c$ of the AA stacked bilayer is a commensurate angle of the Bernal stacked bilayer.
The conductance peaks then lie symmetrically with respect to $\theta=30^\circ$ since if $\theta$ is commensurate so is its inverse.

\section{Conductance for commensurate angles for $\bm{d}=0$.    \label{sec_Gcommensurate}}

For commensurate angles the conductance can be approximated by Eq.(4) in the main text. The integration over the overlap of the two spectral functions
\be
\sum_{\bm{k}} A_1(k,\epsilon_F) A_1(k,\epsilon_F) = \frac{\cA}{2\pi v^2} \left[ 2 + 2\pi\epsilon_F \tau + 4\epsilon_F\tau \arctan\lp 2\epsilon_F\tau \rp \right]
\ee
is independent of the rotation angle and the entire dependence of $G$ on the relative alignment of the two layers is in $R_{\mu\nu}(\theta,\bm{d})$. We now evaluate
$R$ for $\bm{d}=0$.

At the Dirac point the intra-layer Hamiltonian vanishes and we have contributions only from
interlayer tunneling:
\be
H_0 =
\left(
  \begin{array}{cc}
    0 & \cT \\
    \cT^{\dagger} & 0 \\
  \end{array}
\right)        \label{H0}
\ee
where each element is a $2 \times 2$ block for the two $\pi$-bands in each layer.
Using a representation of sublattice sites in each layer we find that
\be
\cT=\cT^{\ty S}=
\left(
  \begin{array}{cc}
    2|a| & 0 \\
    0 & 2|a| \cdot e^{-2i\phi} \\
  \end{array}
\right)
\ \ , \ \ \cT=\cT^{\ty D}=
\left(
  \begin{array}{cc}
    2|a| & 0 \\
    0 & 0 \\
  \end{array}
\right).        \label{Tlattice}
\ee
Here $\cT^{\ty S}$ and $\cT^{\ty D}$ correspond respectively to intra-valley (S=same) and inter-valley (D=different) rotation angles
as explained in the main text, and $\phi=0,\pm 60$ depends on $\theta_c$: {\em e.g.} $\phi(0^\circ)=0$, $\phi(27.8^\circ)=60^\circ$ and $\phi(38.2^\circ)=-60^\circ$.
If the hopping amplitude $t_{q}$ decreases fast enough with momentum so that only the first $G$-shell significantly contribute to the tunneling matrix
\be
|a|= 1.5 \frac{t_{\bm{k_D+G_1}}}{\Omega_0}      \label{a}
\ee
where $\Omega_0$ is the area of a unit cell and $\bm{G}_{1}$ is the
wavevector which produces the smallest $\bm{q}$ extended-zone Dirac-cone overlap as explained in the text.
In our model Eq.(\ref{a}) is satisfied for all commensurate angles except for $\theta=0^\circ,60^\circ$ for which $|a|= 1.67 t_{\bm{k_D+G_1}}/\Omega_0$.
Diagonalizing $H_0$ yields $E^{\ty S} = \pm 2 |a|$ (both doubly degenerate) and $E^{\ty D} = 0,0,\pm 2|a|$.  In both cases the energy gap between the top conduction band
and bottom valence band is therefore $E_g=4|a|$.

To find $R$ we assume that $T$ is well approximated by Eq.(\ref{Tlattice}) for finite momentum states in the vicinity of the Dirac points.
We verified this assumption numerically for low densities. We first focus on inter-valley rotation angles.
In the eigenstate representation
\be
T^{\mu\nu}_{kp'} = a_{k\mu}^{1\alpha\star} a_{k\nu}^{1\beta} \cT^{\alpha\beta}_{kp'} \delta_{kp'}.       \label{lattice2eigenstate}
\ee
It follows from Eqs.(\ref{Tlattice},\ref{lattice2eigenstate}) that
\be
T^{\ty D} = e^{i(2\theta_k+\theta_c)} |a| I.     \label{interK}
\ee
Consequently,
\be
R^{D}_{\mu\nu} = \int \frac{d\theta_k}{2\pi} |T^{\mu\nu}(\theta_k)|^2 = |a| I
\ee
and
\be
G^{\ty D}(\theta_c,d=0) = \cA g_v g_s \frac{e^2}{\hbar} \frac{E^2_g(\theta_c)}{64\pi^2 v^2}\left[ 2 + 2\pi\epsilon_F \tau + 4\epsilon_F\tau \arctan\lp 2\epsilon_F\tau \rp \right].    \label{GD}
\ee
Expression (5) is obtained in the $\epsilon_{\ty F}\tau > 1$ limit.

Similarly for intra-valley rotation angles
\be
T^{\ty S} = 2 e^{i(\phi+\frac{\theta_c}{2})}|a|
\left(
  \begin{array}{cc}
    \cos\lp \phi - \frac{\theta_c}{2} \rp & -i\sin\lp \phi - \frac{\theta_c}{2} \rp \\
    -i\sin\lp \phi - \frac{\theta_c}{2} \rp & \cos\lp \phi - \frac{\theta_c}{2} \rp \\
  \end{array}
\right)
\ee
It then follows that
\be
R^{\ty S}_{\mu\nu} = \frac{E_g^2}{4}
\left(
  \begin{array}{cc}
    \cos^2\lp \phi - \frac{\theta_c}{2} \rp & \sin^2\lp \phi - \frac{\theta_c}{2} \rp \\
    \sin^2\lp \phi - \frac{\theta_c}{2} \rp & \cos^2\lp \phi - \frac{\theta_c}{2} \rp \\
  \end{array}
\right)
\ee
and that the conductance is
\be
G^{\ty S}(\theta_c,d=0) =  \cA g_v g_s \frac{e^2}{\hbar} \frac{E^2_g(\theta_c)}{16\pi^2 v^2} \cos^2\lp \phi - \frac{\theta_c}{2} \rp
\left[ 2 + 2\pi\epsilon_F \tau + 4\epsilon_F\tau \arctan\lp 2\epsilon_F\tau \rp \right].      \label{GS}
\ee

Note that inter-band resonant conduction (which occurs when the carrier densities in the two layers are opposite)
has the same form as its intra-band counterpart for
inter-valley rotation angles.
For the inter-band conduction at inter-valley rotation angles
the $\cos$ function in Eq.(\ref{GS}) should be replaced by a $\sin$. Interestingly, the ratio
\be
\Delta G(\theta_c) \equiv \frac{G^{\ty S}(\theta_c,d=0) }{G^{\ty D}(60^\circ-\theta_c,d=0) } = 4 \cos^2\lp \phi - \frac{\theta_c}{2} \rp
\ee
depends only on the twist angle. For example, $\Delta G(27.8^\circ)=1.94$ in accord with the numerical results depicted in Fig.\ref{fig:Gd}.

Using the momentum dependent $T$ matrices we can find the bands in the vicinity of the Dirac points.
For inter-valley rotations we find four non-degenerate bands
\be
E^{\ty D}_k = \pm \sqrt{\epsilon_k^2 + 2|a|^2 \pm 2\sqrt{\epsilon_k^2 |a|^2 + |a|^4}}.
\ee
At low energies $\epsilon_k \ll a $
\be
E^{\ty D}_{k1} = \pm \frac{k^2}{2m^\star}  \ \ \ , \ \ \ E^{\ty D}_{k2} = \pm 2|a| \pm \frac{k^2}{2m^\star}
\ee
where $m^\star = |a|/v^2$.
For the intra-layer rotations
\be
E^{\ty S}_k = \pm \sqrt{\epsilon_k^2+ 4|a|^2 \pm 4|a|\epsilon_k \cos\lp \phi - \theta_c/2 \rp}
\ee
At low energies $\epsilon_k \ll a $
\be
E^{\ty S}_k = \pm 2|a| \pm v^\star k
\ee
where $ v^\star = v \cos\lp \phi - \theta_c/2 \rp $. Deviations from expressions (\ref{Tlattice}) for $\cT$ result in trigonal warping in a bilayer system.
More elaborate studies of the spectrum are needed to determine whether such effects are important in a rotated bilayer system as well.

\end{widetext}

\end{document}